\pdfoutput=1
\documentclass{ifacconf}

\makeatletter
\let\old@ssect\@ssect 
\makeatother

\usepackage{natbib}
\usepackage{hyperref}

\makeatletter
\def\@ssect#1#2#3#4#5#6{%
  \NR@gettitle{#6}
  \old@ssect{#1}{#2}{#3}{#4}{#5}{#6}
}
\makeatother

\usepackage{graphicx}      
\usepackage{natbib}        

\usepackage{amsmath,amssymb,amsfonts}
\usepackage{algorithmic}
\usepackage{graphicx}
\usepackage{textcomp}
\usepackage{siunitx}
\usepackage{subcaption}
\usepackage{mathtools} 
\usepackage{rotating}
\usepackage{tikz}
\usepackage{pgfplots}
\usepgfplotslibrary{statistics}
\usetikzlibrary{patterns}
\usepackage{tkz-euclide}
\usepackage{tikz}
\usetikzlibrary{shapes.geometric,positioning,intersections,shapes,decorations, arrows.meta, calc, plotmarks, patterns,angles,quotes}
\tikzstyle{block} = [rectangle, draw, rounded corners, text centered, text width = 7em, minimum height = 2em]
\tikzstyle{pfeil} = [draw, -latex']
\tikzstyle{linie} = [draw, --]
\tikzstyle{sum} = [circle, draw, radius=1pt, color=black, fill=white, inner sep=2pt]
\tikzstyle{knoten} = [circle, draw, radius=1pt, color=black, fill=black, inner sep=2pt]

\newboolean{showFigures}
\setboolean{showFigures}{true}
\definecolor{tublau}{RGB}{0,95,140}	
\definecolor{eitcyan}{RGB}{78,188,206}
\definecolor{archblack}{RGB}{0,0,0}
\definecolor{bauorange}{RGB}{255,89,0}
\definecolor{biogruen}{RGB}{1,147,119}
\definecolor{chemorange}{RGB}{230,146,2}
\definecolor{inflila}{RGB}{71,0,116}
\definecolor{maschblau}{RGB}{0,160,198}
\definecolor{mathgelb}{RGB}{254,192,0}
\definecolor{phyblau}{RGB}{7,43,97}
\definecolor{planorange}{RGB}{255,140,0}
\definecolor{sozgrau}{RGB}{82,87,89}
\definecolor{wiwirot}{RGB}{179,0,13}
\DeclareMathOperator*{\diag}{diag} 
\newcommand{\matvar}[1]{\boldsymbol{#1}}
\newcommand{\refEquation}[1]{(\ref{#1})}

\newcommand{\refSection}[1]{Sec.~\ref{#1}}

\newcommand{\refFigure}[1]{Fig.~\ref{#1}}

\def\BibTeX{{\rm B\kern-.05em{\sc i\kern-.025em b}\kern-.08em
    T\kern-.1667em\lower.7ex\hbox{E}\kern-.125emX}}

\pgfplotsset{compat=1.18}
\pdfoutput=1
\begin{document}

\begin{frontmatter}

\title{Trajectory Planning with Model Predictive Control for Obstacle Avoidance Considering Prediction Uncertainty} 
\thanks[footnoteinfo]{The authors gratefully acknowledge the financial support by the German Federal Ministry of Education and Research in the project Open6GHub (grant number: 16KISK004).}
\author[First]{Eric Schöneberg} 
\author[Second]{Michael Schröder} 
\author[First]{Daniel Görges}
\author[Second]{Hans D. Schotten}
\address[First]{Institute of Electromobility, 
   RPTU University Kaiserslautern-Landau, Kaiserslautern, Germany \\ (e-mail: \{eric.schoeneberg, daniel.goerges\}@rptu.de)}
\address[Second]{Institute of Wireless Communications and Navigation, 
   RPTU University Kaiserslautern-Landau, Kaiserslautern, Germany \\(e-mail: \{michael.schroeder, hans.schotten\}@rptu.de)}
\begin{abstract}  
This paper introduces a novel trajectory planner for autonomous robots, specifically designed to enhance navigation by incorporating dynamic obstacle avoidance within the Robot Operating System 2 (ROS2) and Navigation 2 (Nav2) framework. The proposed method utilizes Model Predictive Control (MPC) with a focus on handling the uncertainties associated with the movement prediction of dynamic obstacles. Unlike existing Nav2 trajectory planners which primarily deal with static obstacles or react to the current position of dynamic obstacles, this planner predicts future obstacle positions using a stochastic Vector Auto-Regressive Model (VAR). The obstacles' future positions are represented by probability distributions, and collision avoidance is achieved through constraints based on the Mahalanobis distance, ensuring the robot avoids regions where obstacles are likely to be. This approach considers the robot's kinodynamic constraints, enabling it to track a reference path while adapting to real-time changes in the environment. The paper details the implementation, including obstacle prediction, tracking, and the construction of feasible sets for MPC. Simulation results in a Gazebo environment demonstrate the effectiveness of this method in scenarios where robots must navigate around each other, showing improved collision avoidance capabilities.
\end{abstract}
\begin{keyword}
Mobile robots, Guidance navigation and control, Multi-agent systems, Nonlinear predictive control, Trajectory and Path Planning, Obstacle Avoidance
\end{keyword}
\end{frontmatter}

\begin{tikzpicture}[remember picture,overlay]
    \node[anchor=north,yshift=-10pt] at (current page.north) {\shortstack{\\ \\
		  \fbox{\parbox{\dimexpr\textwidth-\fboxsep-\fboxrule\relax}{\textcopyright 2025 the authors. This work has been accepted to IFAC for publication under a Creative Commons Licence CC-BY-NC-ND.}}} };
\end{tikzpicture}%

\section{Introduction}\label{sec: Introduction}
\begin{figure}[ht]
\includegraphics[width=0.45\textwidth, bb=0 0 800 600]{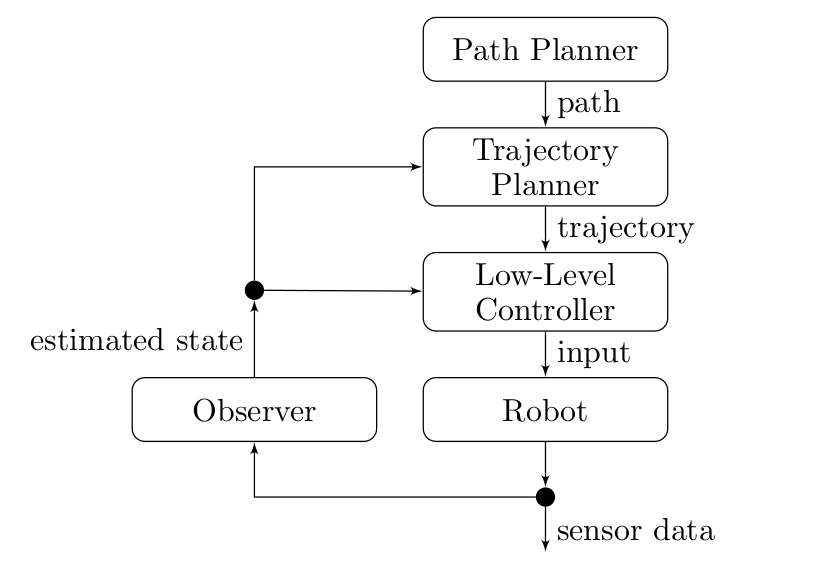}
\caption{Overview: Control Structure}\label{fig:Overview: Control Structure}
\end{figure}

Navigation is crucial for autonomous robots, enabling them to move safely and efficiently in various environments. The established control architecture for this problem is a cascade of path planner, trajectory planner and a low-level controller. The planned path is a reference to the trajectory planner and the trajectory is the reference to the low-level controller as referenced in \refFigure{fig:Overview: Control Structure}. The path planner finds a path (an ordered list of positions) in the highly non-convex task space, but is usually agnostic to the robots kinematics or dynamics and can have a low sample-rate. A trajectory planner finds a trajectory (a chronological list of states) in the robot's configuration space. It accounts for the robot's kinodynamics and tracks the path. A low-level controller then tracks the trajectory. \\
In this scenario, the trajectory planner can account for collision avoidance of dynamic obstacles. Static obstacles can be taken into account during the path planning phase. However, the planned path is not a function of time, hence the trajectory planner must handle collision avoidance for dynamic obstacles. An obstacle is dynamic if it moves relative to a world-fixed frame. To avoid dynamic obstacles over time, their path can be predicted and then considered within the trajectory planning. In this paper, a trajectory planner for path tracking with dynamic obstacle avoidance based on Model Predictive Control (MPC) is presented. The approach does not rely on communication between agents and accounts for the estimated uncertainties of the obstacle motion prediction model. While any prediction method that estimates future expected values and the associated covariance matrices are compatible with the  approach, a stochastic Vector Auto-Regressive Model (VAR) \cite[pp. 69--80]{lutkepohl2005new} is proposed. 
Robot Operating System 2 (ROS2) \cite{ros2} is the de facto standard in robotic applications for both academics and the maturing robotics industry. Within this ecosystem, Navigation 2 (Nav2) \cite{nav2} stack is a comprehensive framework designed to support robot navigation, including tasks such as path planning, obstacle avoidance, and localization. The goal of this paper is to introduce a trajectory planner compatible within this established environment, that has proven to work in both simulation and real-world applications and that can be tested within an accessible framework. As illustrated in Fig.~\ref{fig:Overview: Control Structure}, the overall Nav2 control structure is cascaded.
Different controllers operate asynchronously, with the path planner typically sampling at a lower rate and the low-level controller sampling at a higher rate than the trajectory planner.
\subsection{Related Work}\label{sec: Related Work}
Two aspects of related work regarding trajectory planning are discussed: practical considerations using Nav2 and theoretical approaches without Nav2 integration, yet relevant to the presented method.
Nav2 provides a variety of trajectory planners: Dynamic Window Approach (DWA/DWB) \citep{DWA}, Vector Pursuit Controller (VPC) \citep{wit2000vector}, Graceful Controller (GC) \citep{park2011smooth}, Model Predictive Path Integral (MPPI) \citep{mpii}, Timed Elastic Band (TEB) \citep{rosmann2017integrated} and Regulated Pure Pursuit (RPP) \citep{macenski2023regulated}. \\
DWA/DWB considers static obstacles and calculates trajectories in velocity space, only allowing for velocities in a set defined by accessible accelerations. To guarantee collision avoidance, it only allows for trajectories where a safe stop is possible -- given obstacles can be considered static at the time of perception, meaning obstacles are treated as stationary at the moment they are detected, without accounting for their future motion. VPC calculates a trajectory based on screw theory to follow a predetermined path without considering kinematic constraints specifically. GC is a non-linear control law to smoothly follow a path, assuming the robot can be approximated by a simple unicycle model. It is proven to be globally exponentially stable. MPPI considers a stochastic dynamical system which is affine to the control input and shows that there exists an iterative update rule to find an optimal sequence of inputs. It handles static and dynamic obstacles by penalizing trajectories entering regions of collision. The first entry of this sequence is then implemented in a Receding Horizon fashion. Neither DWA, VPC, nor GC handle obstacle avoidance directly. Instead, they follow predefined paths, with obstacle avoidance managed by the path planner. All obstacles are considered static by all planners. TEB proposes an MPC, which employs an online optimization of constrained nonlinear quadratic problems, of which one is chosen. It examines obstacle representations through sets of points, circles, lines, and polygons. The emphasis is to find the dual problem formulation which is conducive to the application. The problem solver can then rapidly converge to a solution with reduced computational burden. Obstacle avoidance is handled by an inequality constraint. The robot is represented by a circle and any obstacle as a simply-connected region in $\mathbb{R}^2$. The inequality constraint denotes a minimal Euclidean distance between robot and region. TEB considers obstacles to be static at the time of perception over the prediction horizon. RPP uses simple geometry to find a suitable curvature to follow a given path. It considers obstacles by a proximity heuristic, that limits velocity in close proximity to dynamic or static obstacles, while considering all obstacles static at the time of perception. \\
An approach beyond those available in Nav2 by \cite{Hardy2013} introduces the description of obstacle positions with a probability distribution at any point of time in the future in the context of trajectory planning. The authors then propose an upper bound on the collision probability between robot and obstacle, which is considered a constraint in a dynamic program. The collision probability in this case can be calculated with an arbitrary polygonal robot shape. \cite{Bhatt2019} propose a stochastic Model Predictive Controller with constraints on the ellipse overlaps of robot and obstacle probability distribution based on the Bhattacharyya distance \citep{bhattacharyya1943measure}, which is then introduced as a constraint. \cite{Zhang2021} propose motion prediction based on machine-learning Mixture Density Networks resulting in Gaussian probability distributions of obstacles over time. The prediction would be used as a constraint based on the Mahalanobis distance in a Model Predictive Controller. Yet the authors do not provide context on how the model predictive controller would integrate the information or implementation of it, but rather focus on prediction. \\
Firstly, it can be argued that trajectory planning is significantly enhanced when considering obstacles as dynamic. This consideration allows for predictive acting to the changing positions of other entities, instead of reacting to current and past positions. Secondly, predicting the future movement of obstacles must inherently be uncertain -- especially in scenarios where agents do not communicate. The future positions of obstacles depend on multiple uncertain factors, including the agents' unknown goals, their limited and potentially inaccurate perception, and the semantic context of the environment, which influences movement patterns. Third, a trajectory planner needs capabilities to follow a given path, respect the robot's kinodynamic limitations, plan over time and act on changes in the environment which the path planner does not consider. The choice of a Model Predictive Controller is apparent. 
\subsection{Contribution}
The proposed trajectory planner has the advantage and novelty over existing Nav2 controllers to consider dynamic obstacles. Unlike the previously discussed approaches, the focus is on the construction of the MPC strategy and especially on the construction of constraints based on prediction uncertainties. To illustrate its behavior and properties, the trajectory planner is implemented in a well-established robotics setting, facilitating testing, benchmarking, and hardware implementation with minimal effort. The source code for the proposed trajectory planner will be released as open-source upon achieving feature completeness and stability, incorporating obstacle avoidance for static polygonal obstacles and tracking of dynamic obstacles. No modifications to the Nav2 stack are required for integration, ensuring straightforward adoption. The planner’s architecture supports any motion prediction for dynamic obstacles as a time series with normally distributed noise, exemplified here by a VAR process, providing a robust foundation for future motion prediction research within a fully functional technology stack. Using Casadi for problem modeling, the planner accommodates any robot kinematics, with a Differential Drive implementation provided as an example. The architecture scales seamlessly to multiple dynamic obstacles, with performance tied to the integration of obstacle tracking and the corresponding numbers of constraints.
\section{Model Predictive Control}\label{sec:MPC}
Consider a robot with state $\matvar{x} \in \mathbb{R}^n$, input $\matvar{u} \in \mathbb{R}^m$ and some non-linear discrete-time dynamics $\matvar{x}_{k+1} = f\left( \matvar{x}_{k}, \matvar{u}_{k} \right)$ where $f: \mathbb{R}^n \times \mathbb{R}^m \rightarrow \mathbb{R}^n$ and $k\geq \mathbb{N}_{>0}$ is the current discrete time. Over some prediction horizon $N \in \mathbb{N}_{>0}$ define the state and input trajectory $\matvar{X}_k = \begin{pmatrix} \matvar{x}_{k+1}^\intercal & \matvar{x}_{k+2}^\intercal \dots & \matvar{x}_{k+N}^\intercal \end{pmatrix}^\intercal \in \mathbb{R}^{Nn}$ and $\matvar{U}_k = \begin{pmatrix} \matvar{u}_{k}^\intercal & \matvar{u}_{k+1}^\intercal \dots & \matvar{u}_{k+N-1}^\intercal \end{pmatrix}^\intercal \in \mathbb{R}^{Nm}$. Then the general MPC problem for trajectory planning can be written as
\begin{subequations}\label{eq:MPCproblem.1}
    \begin{align}
        &\underset{\left\{\matvar{U}_k, \matvar{X}_k \right\}}{\text{minimize }}   && f_0 \left(\matvar{X}_k , \matvar{U}_k\right) \label{eq:MPCproblem.1a}\\
        &\text{subject to } && \matvar{x}_{k+i+1} = f\left(\matvar{x}_{k+i}, \matvar{u}_{k+i} \right)  \label{eq:MPCproblem.1b}\\
        &~                  && \matvar{x}_{k+i} \in \mathbb{X}_{k+i} \label{eq:MPCproblem.1c} \\
        &~                  && \matvar{u}_{k+i-1} \in \mathbb{U}_{k+i-1} \label{eq:MPCproblem.1d}
    \end{align}
\end{subequations}
where $f_0: \mathbb{R}^{Nn}\times\mathbb{R}^{Nm} \rightarrow \mathbb{R}$ is differentiable, $\mathbb{X}_{k+i+1} \subseteq \mathbb{R}^{n}$ and $\mathbb{U}_{k+i-1} \subseteq \mathbb{R}^{m}$ $\forall i = 1, \dots, N$. This non-convex optimization problem will be solved for $\left\{\matvar{U}_{\text{r},k}, \matvar{X}_{\text{r}, k} \right\}$ in a Receding Horizon fashion. $\matvar{X}_{\text{r},k}$ is the desired trajectory. The focus here is on dynamic obstacle avoidance in the two-dimensional robot workspace $\mathbb{W}\subset\mathbb{R}^2$, where only the position $\matvar{p}_{k+i} = \begin{pmatrix} x_{k+i} & y_{k+i} \end{pmatrix}^\intercal \in \mathbb{W}$ where $\matvar{p}_{k+i}$ as a part of state $\matvar{x}_{k+i} = \begin{pmatrix} \dots & \matvar{p}_{k+i} & \dots \end{pmatrix}^\intercal \in \mathbb{R}^{n}$ will be restricted. If the problem includes obstacle avoidance, the feasible set is 
\begin{equation}\label{eq:set_construction}
    \mathbb{X}_{k+i+1} = \left\{\matvar{x}_{k+i} \in \mathbb{R}^n \left| \matvar{p}_{k+i} \in \left\{\mathbb{S}^\complement \cap \mathbb{D}_{k+i}^\complement\right\} \right. \right\}
\end{equation} 
where $\mathbb{S}$ describes the set of static obstacles and $\mathbb{D}_{k+i}$ describes the set of dynamic obstacles at time $k+i$. Three-dimensional obstacle avoidance works in the same way. The construction of the sets will be discussed in \refSection{sec: Set Construction}.
\section{Set Construction}\label{sec: Set Construction}
\begin{figure}
 \includegraphics[width=\linewidth, bb=0 0 900 600]{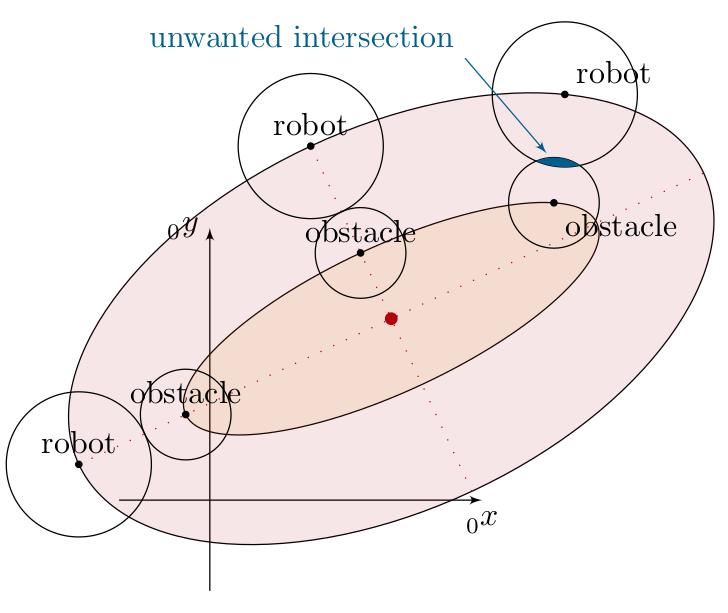}
\caption{Approximation Error in Ellipse Construction}
\label{fig:approximation error}
\end{figure}

Any dynamic obstacle $d\in\{1,\ldots,o\}$ where $o\in\mathbb{N}$ has a radius $r_d$ and position $\matvar{D}_{d,k+i}$ at time $k+i$, where  $\matvar{D}_{d,k+i}\sim\mathcal{N}\left(\matvar{\mu}_{d, k+i}, \matvar{\Sigma}_{d,k+i}\right)$ and $\mathcal{N}$ denotes the normal distribution with $\matvar{\mu}$ and $\matvar{\Sigma}$ the expectation and covariance respectively. The Mahalanobis distance \citep{mahalanobis1936generalized} $d_M$ of a point $\matvar{x}$ to a normal distribution $\mathcal{N}\left(\matvar{\mu}, \matvar{\Sigma}\right)$ is defined as
\begin{equation}\label{eq:mahalanobis distance}
    d_M^2 \left(\matvar{x}; \mathcal{N}\left(\matvar{\mu}, \matvar{\Sigma}\right)\right) = \left(\matvar{x} - \matvar{\mu}\right)^\intercal \matvar{\Sigma}^{-1}\left(\matvar{x} - \matvar{\mu}\right) 
\end{equation}
In a first step, $\mathbb{D}$ should describe the set in which the dynamic obstacle will be with a given probability $p$. This can be achieved by
\begin{equation}\label{eq:dynamic_sets}
    \mathbb{D} = \left\{ \matvar{x} \in \mathbb{R}^n | d_M < s  \right\} 
\end{equation}
Since $d_M^2$ is $\chi^2_2$-distributed, the threshold has to be chosen as $s = \sqrt{-2\ln{\left(1-p\right)}}$ leading to the \textcolor{chemorange}{inner ellipse} in \refFigure{fig:ellipse}. For 3-dimensional obstacles, $s$ would be chosen according to the $\chi^2_3$-distribution. For collision avoidance, this is not sufficient yet. The intersection of the robot with any of the obstacles has to be avoided. The robot and the obstacle $d$ are conservatively approximated by a circle with radius $r$ and $r_d$ respectively. Enlarging the ellipse by $r+r_d$ is easier after the eigen-decomposition of $\matvar{\Sigma}$, i.e.
\begin{subequations}\label{eq:mahalanobis distance decomposition}
\begin{align}
    d_M^2 \left(\matvar{x}; \mathcal{N}\left(\matvar{\mu}, \matvar{\Sigma}\right)\right) &= \left(\matvar{x} - \matvar{\mu}\right)^\intercal \matvar{R}\matvar{\Lambda}^{-1}\matvar{R}^{-1}\left(\matvar{x} - \matvar{\mu}\right) \label{eq:mahalanobis distance.b}\\
    &= \left(\matvar{x} - \matvar{\mu}\right)^\intercal \matvar{R}\matvar{\Lambda}^{-1}\matvar{R}^\intercal\left(\matvar{x} - \matvar{\mu}\right) \label{eq:mahalanobis distance.c}
\end{align}
\end{subequations}
where $\matvar{\Lambda}$ is a diagonal matrix with $\matvar{\Sigma}$'s eigenvalues $\lambda_1$ and $\lambda_2$ as its diagonal entries. $\matvar{R}$'s columns are $\matvar{\Sigma}$'s eigenvectors $\matvar{v}_1$ and $\matvar{v}_2$ with length $1$ and $\matvar{R}^\intercal = \matvar{R}^{-1}$, since $\matvar{R}$ is orthogonal. Using the coordinate transformation $\matvar{x}^\prime = \matvar{R}^\intercal \left(\matvar{x}-\matvar{\mu}\right)$ and \refEquation{eq:mahalanobis distance decomposition}, $d_M^2 < s^2$ can be expressed based on \refEquation{eq:dynamic_sets} as
\begin{subequations}
\begin{align}
    \matvar{x}^{\prime\intercal} \matvar{\Lambda}^{-1} \matvar{x}^\prime &< s^2 \label{eq:scaled_dynamic_set.a}\\
\left(\frac{x^\prime_1}{s\sqrt{\lambda_1}}\right)^2 + \left(\frac{x^\prime_2}{s\sqrt{\lambda_2}}\right)^2 &< 1 \label{eq:scaled_dynamic_set.b}
\end{align}
\end{subequations}
to describe the ellipse in a coordinate frame which axes align with the axes of the ellipse and whose origin is in $\matvar{\mu}$. Equation \refEquation{eq:scaled_dynamic_set.b} shows that the ellipse's axes have the lengths $s\sqrt{\lambda_1}$ and $s\sqrt{\lambda_2}$, which can be enlarged by redefining $\matvar{\Lambda}^{-1}$
as
\begin{subequations}\label{eq:ellipse parts}
\begin{align}
        \matvar{\Lambda}_{d,k+i}^{-1} &=  \diag \begin{pmatrix}1/\left(s\sqrt{\lambda_{d,1,k+i}}+r + r_d\right)^2 \\ 1/\left(s\sqrt{\lambda_{d,2,,k+i}}+r + r_d\right)^2\end{pmatrix} \label{eq:inverse lambda}
        \intertext{with}
        \matvar{R}_{d,k+i} &= \begin{bmatrix}
            \matvar{v}_{d,1,k+i} & \matvar{v}_{d,2,k+i} 
        \end{bmatrix} \label{eq:eigenvector matrix}\\
        \Delta\matvar{p}_{d,k+i} &= \matvar{p}_{k+i} - \matvar{\mu}_{d,k+i} \label{eq:delta p}
\end{align}
\end{subequations}
and \refEquation{eq:mahalanobis distance.c}. This leads to
\begin{equation}\label{eq:ellipse set}
\begin{aligned}
    \mathbb{D}_{d,k+i} = \left\{ \matvar{p}_{k+i} \in \mathbb{R}^2 | \Delta\matvar{p}_{d,k+i}^\intercal \matvar{R}_{d,k+i} \right.&\\
      \left.\matvar{\Lambda}_{d,k+i}^{-1} \matvar{R}_{d,k+i}^\intercal \Delta\matvar{p}_{d,k+i} < 1  \right\}&
  \end{aligned}
\end{equation}
\begin{equation}\label{eq:compliment ellipse set}
\begin{aligned}
    \mathbb{D}_{d,k+i}^\complement = \left\{ \matvar{p}_{k+i} \in \mathbb{R}^2 | \Delta\matvar{p}_{d,k+i}^\intercal \matvar{R}_{d,k+i} \right.&\\
      \left.\matvar{\Lambda}_{d,k+i}^{-1} \matvar{R}_{d,k+i}^\intercal \Delta\matvar{p}_{d,k+i} \geq 1  \right\}&
  \end{aligned}
\end{equation}
which is shown as the \textcolor{wiwirot}{outer ellipse} in \refFigure{fig:ellipse}.
\ifthenelse{\boolean{showFigures}}{\begin{figure}[ht]
\includegraphics[width=\linewidth, bb=0 0 1000 800]{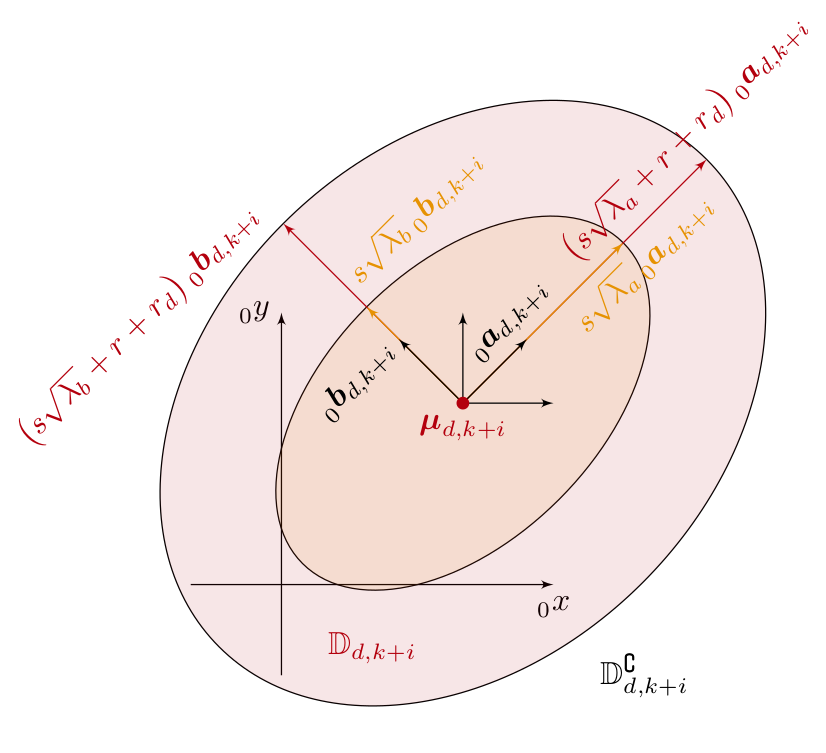}
\caption{Visual Representation of Ellipse Construction}\label{fig:ellipse}
\end{figure}
}\\
The set in \refEquation{eq:set_construction} to avoid all dynamic obstacles $o$ is therefore $\mathbb{D}_{k+i}^\complement = \bigcap_d \mathbb{D}_{d,k+i}^\complement$ combined at every time-step $k+i$.
\begin{figure}
        \includegraphics[width=\linewidth, bb=0 0 600 400]{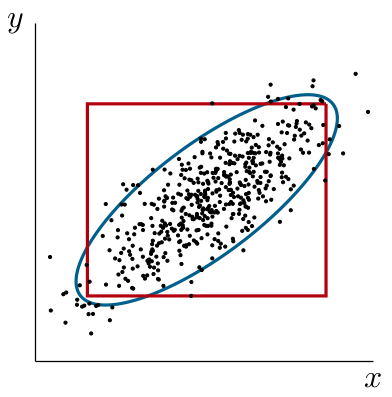}
    \caption{Comparison of different forecast regions for $p=0.90$}
    \label{fig:comparison_forecast_regions}
\end{figure}
The extension of the ellipse by $r+r_d$ is an approximation. As shown in \refFigure{fig:approximation error}, there exists no approximation error if the obstacle's position is along the axis of the ellipse and for perfectly conditioned covariance matrices -- i.e. when the ellipse becomes a perfect circle. Ill-conditioned covariance matrices and deviation from minor and major however result in an approximation error. This imperfection can be compensated by choosing larger $s$ or by expanding the ellipse by more than $r+r_d$.
An alternative way to construct forecast regions is described by \cite[p.~39-41]{lutkepohl2005new} and leads to a rectangle in the 2-dimensional case. Establishing a threshold for the Mahalanobis distance results in an ellipse, which covers a smaller area. The ellipse is therefore the more accurate representation of the underlying distribution and less conservative -- see \refFigure{fig:comparison_forecast_regions}.
\section{Implementation}\label{sec:Implementation}
\begin{figure}[ht]
 \includegraphics[width=\linewidth, bb=0 0 900 750]{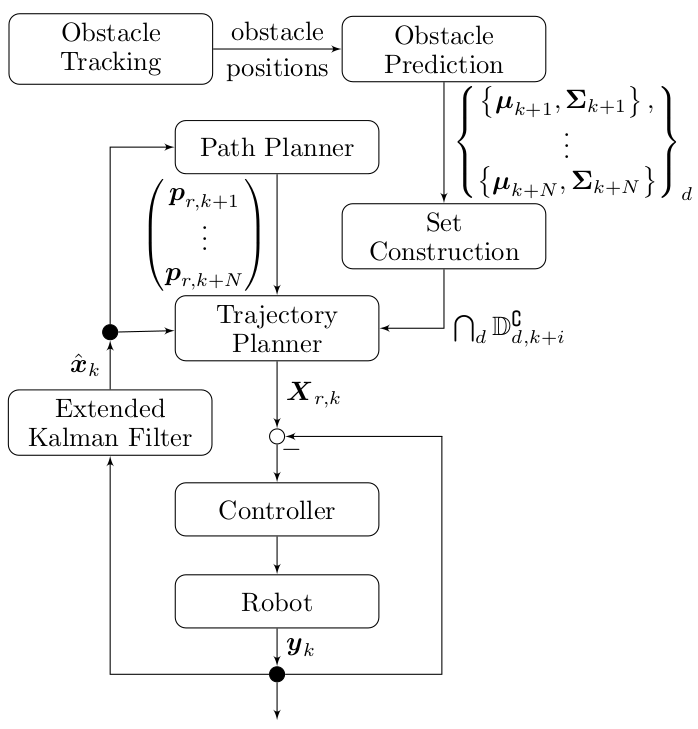}
\caption{Details: Control Structure}\label{fig:Control Structure}
\end{figure}

For the control structure in \refFigure{fig:Overview: Control Structure}, the implementation is discussed here in a detailed fashion following \refFigure{fig:Control Structure}. The simulation presented in the following is for two skid-steered robots equipped with a LiDAR, wheel encoders and IMU in Gazebo \citep{koenig2004design}. The visualization of the robots, paths and trajectories -- as seen in Figs.~\ref{fig:Chicken Game},~\ref{fig:collision avoidance} -- is done in RViz2 \citep{kam2015rviz}. \\
In order to predict obstacle movement, one must first be able to perceive and track current position and possibly heading of dynamic obstacles. This will involve extracting the obstacle from sensor data or from a dynamic map. Once perceived, an obstacle must be tracked consistently over time. Therefore one must solve a matching problem: a list of obstacles is perceived at time $k-1$ and $k$ and one has to know which entries of both list correspond to the same obstacle. Current discussion involves solving this matching problem by applying the Hungarian Algorithm (\cite{navigation2_dynamic_issues, przybyla2017detection}). Obstacle perception and tracking is necessary for a real application, but will be omitted here. To mimic these tasks, communication of current position between robots within the simulation is allowed. \\
Both simulated robots are controlled by the Nav2 stack. They communicate their respective positions and predict each other based on a Vector Auto-Regressive process of order $2$ (VAR(2)) using \textit{statsmodels} \citep{seabold2010statsmodels}. VAR(2) models predict outcomes by summing up linear combinations of previous values. For practicability, the state of the VAR(2) process is therefore defined as the velocity of the obstacle expressed with Cartesian coordinates in a world-fix frame, rather than position. Thus, the expected value and covariance of the velocities ($\matvar{\mu_v}$ and $\matvar{\Sigma_v}$) are estimated instead of the corresponding values for the obstacle's position ($\matvar{\mu}$ and $\matvar{\Sigma}$) which are needed for the MPC. To transform velocities to position, let
\begin{equation}
    \matvar{\mu}_{k+i} = \matvar \mu_{k+i-1} + \Delta T \frac{\matvar{\mu}_{v,k+i} + \matvar{\mu}_{v,k+i-1}}{2}
\end{equation}
and derive
\begin{equation}
    \matvar{\Sigma}_{k+i} = \matvar{\Sigma}_{k+i-1} + \frac{\Delta T^2}{4} \left( \matvar{\Sigma}_{\matvar v,k} + \matvar{\Sigma}_{\matvar v, k+i-1}\right),
\end{equation}
while neglecting the covariances between the summands, $\forall i = 1, \dots, N$.
An Extended Kalman Filter \citep{MooreStouchKeneralizedEkf2014} estimates the state of the robot. The path planner is NavFn planner -- a Dijkstra Algorithm \citep{Dijkstra1959} based path finder. \\
The MPC is implemented using CasADi \citep{Andersson2019} and the resulting program is solved with IPOPT \citep{wachter2006implementation}. The skid-steered robot can be described by non-linear unicycle kinematics 
\begin{align}
    \dot{\matvar x} &= \begin{pmatrix}
        \dot x \\
        \dot y \\
        \dot \phi \\
        \dot v \\
        \dot \omega
    \end{pmatrix} = 
    \begin{pmatrix}
        \cos(\phi) \cdot v \\
        \sin(\phi) \cdot v \\
        \omega \\
        a \\
        \alpha \\
    \end{pmatrix}\label{equ:mpc_modell_kontinuierlich}
\end{align}
where $\matvar{x}=\begin{pmatrix}x & y & \phi & v & \omega \end{pmatrix}^\intercal$ are the robot's position $x$ and $y$, yaw $\phi$, velocity $v$ and yaw rate $\omega$. The system's input $\matvar{u}=\begin{pmatrix}a & \alpha\end{pmatrix}^\intercal$ is translational acceleration $a$ and rotational acceleration $\alpha$. This model is then discretized using symbolic Runge-Kutta of order $4$. The symbolic expression is then used as the equality constraint in \refEquation{eq:MPCproblem.1b}. 
Given an ordered sequence of points as a reference path $\matvar{p}_{\text{r},k+i}= \left(x_{\text{r},k+i}, y_{\text{r},k+i}\right)$ with  $i=1,\dots,N$, the objective function is chosen to be weighted Euclidean distance from path and constant velocity. Define $\matvar{x}_{\text{ref},k+i} = \begin{pmatrix} x_{\text{r},k+i} & x_{\text{r},k+i} & 0 & v_{\text{ref}} & 0\end{pmatrix}^\intercal$, with constant $v_{\text{ref}} = \SI{0.5}{\meter/\second}$. The obstacle avoidance is implemented as probabilistic constraint, by including $(s-s_{\text{ref}})^2$ in the cost function and setting $s_{\text{r}}$ according to the $\chi^2_2$-distribution, while $s$ is another decision variable. Let the prediction horizon $N=15$ with sample rate $\SI{2}{\hertz}$ and $s_{\text{ref}}=\sqrt{5.991}$ for a confidence region of $95\%$. Additionally, restrict translational and rotational velocity $v_{\text{max}} = \SI{0.7}{\meter/\second}$, $\omega_{\text{max}} = \SI{0.3}{\second^{-1}}$ and translational and rotational acceleration $a_{\text{max}} = \SI{0.7}{\meter/\second^2}$, $\alpha_{\text{max}}=\SI{0.1}{\second^{-2}}$ respectively. The MPC problem \refEquation{eq:MPCproblem.1a} to \refEquation{eq:MPCproblem.1d} with sets \refEquation{eq:compliment ellipse set} becomes
\begin{subequations}\label{eq:MPCproblem_implemented.1}
    \begin{align}
        &\underset{\left\{\matvar{U}_k, \matvar{X}_k, s \right\}_{}}{\text{minimize }}   && \sum_{i=1}^{N} \left(\matvar{x}_{k+i} - \matvar{x}_{\text{ref},k+i}\right)^\intercal \matvar{Q} \left(\matvar{x}_{k+i} - \matvar{x}_{\text{ref},k+i}\right) \nonumber\\
        &~ && + \matvar{u}_{k+i-1}^\intercal \matvar{P} \matvar{u}_{k+i-1} + (s-s_{\text{ref}})^2\\
        &\text{subject to } && \matvar{x}_{k+i+1} = f\left(\matvar{x}_{k+i}, \matvar{u}_{k+i} \right)  \tag{\ref{eq:MPCproblem.1b}}\\
        &~ &&\Delta\matvar{p}_{d,k+i}^\intercal \matvar{R}_{d,k+i} \matvar{\Lambda}(s)_{d,k+i}^{-1} \matvar{R}_{d,k+i}^\intercal \Delta\matvar{p}_{d,k+i} \geq 1 \\
        &~                  && \begin{pmatrix}
            -v_{\text{max}} \\
            -\omega_{\text{max}} 
        \end{pmatrix}  \leq \begin{pmatrix}
            v_{k+i} \\
            \omega_{k+i} \\
        \end{pmatrix} \leq  \begin{pmatrix}
            v_{\text{max}} \\
            \omega_{\text{max}} 
        \end{pmatrix} \\
        &~                  && \begin{pmatrix}
            -a_{\text{max}} \\
            -\alpha_{\text{max}} 
        \end{pmatrix}  \leq \matvar{u}_{k+i} \leq  \begin{pmatrix}
            a_{\text{max}} \\
            \alpha_{\text{max}} 
        \end{pmatrix}
    \end{align}
\end{subequations}
with $\matvar{\Lambda}_{d,k+i}^{-1}$ \refEquation{eq:inverse lambda}, $\matvar{R}_{d,k+i}$ \refEquation{eq:eigenvector matrix}, $\Delta\matvar{p}_{d,k+i}$ \refEquation{eq:delta p} and $i=1,\dots,N$. Let $\matvar{Q} = \text{diag}\begin{pmatrix} w_p & w_p & 0 & w_v & 0 \end{pmatrix}^\intercal$ and $\matvar{P} = \text{diag}\begin{pmatrix} w_a & w_\alpha\end{pmatrix}^\intercal$ with $w_p=\SI{100}{\meter^{-2}}$, $w_v=\SI{10}{\second^2/\meter^2}$, $w_a=\SI{1e4}{\second^4/\meter^2}$ and $w_\alpha=\SI{500}{\second^4}$.
The output of the MPC is consequently the desired acceleration, but the output of a Nav2-controller-plugin has to be the desired velocity $\begin{pmatrix} v & \omega \end{pmatrix}^\intercal$. Therefore, instead of $\matvar{u}_{k+0}$ the plugin returns $\begin{pmatrix} v_{k+1} & \omega_{k+1} \end{pmatrix}^\intercal$. A velocity smoother is implemented, that internally calculates the acceleration $\matvar{u}_{k+0}$, based on $\begin{pmatrix} v_{k+1} & \omega_{k+1} \end{pmatrix}^\intercal$ and $\begin{pmatrix} v_{k} & \omega_{k} \end{pmatrix}^\intercal$, which is $\begin{pmatrix} v_{k+1} & \omega_{k+1} \end{pmatrix}^\intercal$ of the previous cycle. This can then be integrated starting from $\begin{pmatrix} v_{k} & \omega_{k} \end{pmatrix}^\intercal$, which is executed with a higher frequency than the MPC. An advantage of this velocity smoother is that it can be modeled easily in an MPC, whereas the default smoother in Nav2's \textit{nav2\_velocity\_smoother} cannot. POPT maximum CPU time is configured to be $\SI{450}{\milli\second}$ to guarantee a trajectory at the defined trajectory planner sample rate of $\SI{2}{\hertz}$. 
\section{Results, Conclusion \& Outlook}\label{sec:conclusion}
\begin{figure}[ht]
    \centering
    \begin{subfigure}[b]{0.49\linewidth}
        \centering
        \includegraphics[width=\textwidth, bb=0 0 520 520]{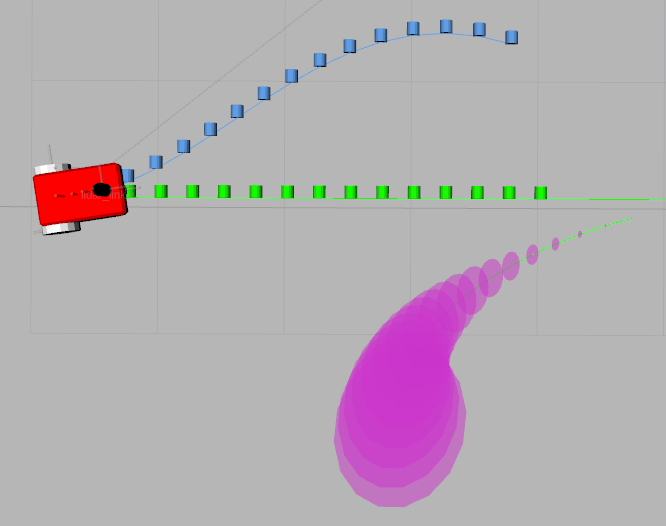}
    \subcaption{Left Robot's view on the Chicken Game}
    \label{fig:Chicken Game Left}
    \end{subfigure}
    \begin{subfigure}[b]{0.49\linewidth}
        \centering
        \includegraphics[width=\textwidth, bb=0 0 520 520]{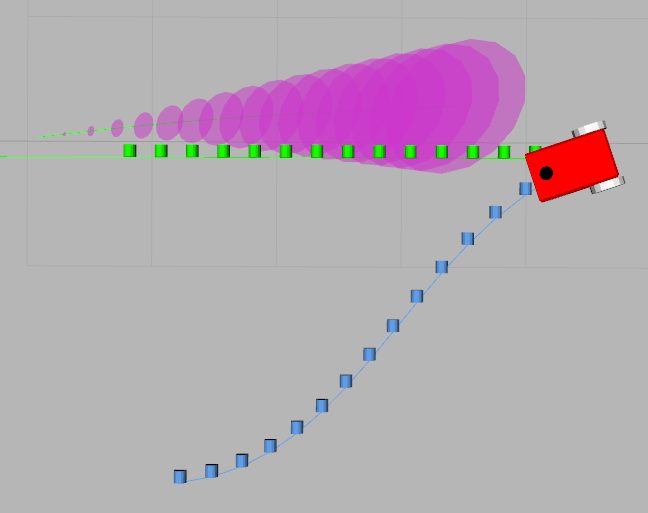}
    \subcaption{Right Robot's view on the Chicken Game}
    \label{fig:Chicken Game right}
    \end{subfigure}
\caption{Two robots in a Chicken Game}
\label{fig:Chicken Game}
\end{figure}

\begin{figure}[ht]
    \centering
    \begin{subfigure}[b]{0.49\linewidth}
    \centering
        \includegraphics[width=0.8\textwidth, bb=0 0 320 320]{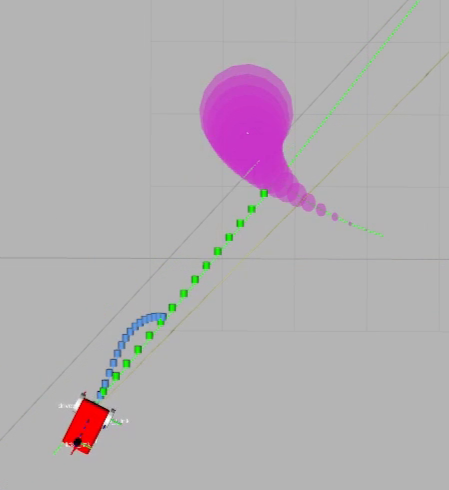}
    \subcaption{Lower Robot's view}
    \label{fig:collision avoidance - lower}
    \end{subfigure}
    \begin{subfigure}[b]{0.49\linewidth}
        \centering
        \includegraphics[width=0.8\textwidth, bb=0 0 320 320]{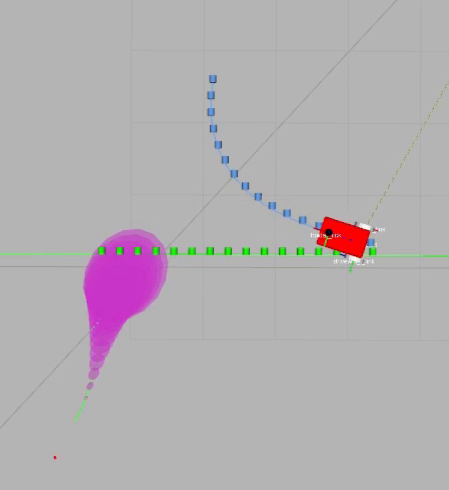}
    \subcaption{Upper Robot's view}
    \label{fig:collision avoidance - upper}
    \end{subfigure}
    \caption{Collision Avoidance}
    \label{fig:collision avoidance}
\end{figure}

Figures \ref{fig:Chicken Game} and \ref{fig:collision avoidance} show examples of two simulated robots avoiding a collision with each other. The green dots represent the reference path, the blue points are the positions of the planned trajectory and the pink areas represent the forecast regions over the prediction horizon before the expansion will enlarge the ellipse as shown in \refFigure{fig:ellipse} by $r + r_d = 2r = \SI{1.5}{\meter}$.\\
The scenario in \refFigure{fig:Chicken Game} depicts two robots in a Chicken or Hawk/Dove Game. Both planned paths predetermine head-on collision. It is resolved by yield of the right robot as perceived by the left robot, see \refFigure{fig:Chicken Game right}. \\
The scenario in \refFigure{fig:collision avoidance} shows two robots avoiding collision where both planned paths cross. Specifically, \refFigure{fig:collision avoidance - lower} shows the advantage of obstacle prediction: the robot sways to the right even though that is the current position of the dynamic obstacle, predicting the opposing robot to move left eventually. \\
Regarding real-time capabilities, solver CPU time typically ranges between \SI{40}{\milli\second} and \SI{50}{\milli\second}, in the tested conditions, where the simulation, Nav2 stack, and trajectory planner are running on an AMD 5800X. Yet, it can spike up to the constraint of \SI{450}{\milli\second}. Given the non-convexity of the optimization problem and the parameter change at each iteration, it is difficult to analyze the spikes generally. Qualitatively, the spikes appear to occur in badly conditioned problems when the situation creates unfavorable parameters. These could include large shifts in the operational environment, such as sudden obstacle movement or changes in surrounding conditions, as well as rerouting or substantial waypoint adjustments by the path planner. It can be noted that untimely trajectory planners are considered and caught in Nav2 in the robot behavior abstraction, hence the solver's performance remains within an acceptable range for real-time operation. Future work can include assessment on real-time capabilities, specifically in comparison with other practical solutions introduced in \refSection{sec: Related Work}. It will be necessary to integrate dynamic obstacles into the Nav2 stack -- specifically to unify the approach to obstacle detection, tracking and interfacing. 
Although this paper does not address the avoidance of static obstacles, incorporation of static obstacle avoidance was already considered in the implementation. Future work will explore the construction of static obstacle sets for use in a MPC strategy and the integration of both dynamic and static obstacle avoidance within one unified framework. With this, future work will include real robot experiments. \\ Generally the subject of overall system characteristics remains for future work. The proposed architecture allows one agent to analyze all other's agents behavior which can be considered in a dynamic game theory approach. It might allow for a model that is symmetric, but still does not rely on communication. 

\bibliography{ifacconf.bib}
\end{document}